\documentclass[11pt,preprint]{aastex}
\usepackage{psfig}  

\slugcomment{Accepted by the Astrophysical Journal}

\shorttitle{Detecting the Cosmic Dipole in Radio Surveys}

\begin{document}

\title{Detecting the Cosmic Dipole Anisotropy in Large-Scale Radio Surveys}

\author{Fronefield Crawford\altaffilmark{1}} 
\altaffiltext{1}{Department of Physics and Astronomy, Franklin \&
Marshall College, Lancaster, PA 17064; email: fcrawfor@fandm.edu}

\begin{abstract}
The detection of a dipole anisotropy in the sky distribution of
sources in large-scale radio surveys can be used to constrain the
magnitude and direction of our local motion with respect to an
isotropically distributed extragalactic radio source population. Such
a population is predicted to be present at cosmological redshifts in
an isotropically expanding universe.  The extragalactic radio source
population is observed to have a median redshift of $z \sim 1$, a much
later epoch than the cosmic microwave background ($z \sim 1100$).  I
consider the detectability of a velocity dipole anisotropy in radio
surveys having a finite number of source counts.  The statistical
significance of a velocity dipole detection from radio source counts
is also discussed in detail.  I find that existing large-scale radio
survey catalogs do not have a sufficient number of sources to detect
the expected velocity dipole with statistical significance, even if
survey masking and flux calibration complications can be completely
eliminated (i.e., if both the surveys and observing instruments are
perfect).  However, a dipole anisotropy should be easily detectable in
future radio surveys planned with next-generation radio facilities,
such as the Low Frequency Array and the Square Kilometer Array; tight
constraints on the dipole magnitude and direction should be possible
if flux calibration problems can be sufficiently minimized or
corrected and contamination from local sources eliminated.
\end{abstract}


\section{Introduction and Motivation}

The measurement of a dipole anisotropy in large-scale surveys can be
used to probe the distribution of matter at different distances and
constrain our local motion with respect to large-scale mass
distributions in the universe. There have been a number of attempts to
measure the dipole anisotropy in different surveys.  The dipole
anisotropy in the cosmic microwave background at redshift $z \sim
1100$, probed by the {\it Cosmic Background Explorer} \citep{sbk+92},
and more recently by the {\it Wilkinson Microwave Anisotropy Probe}
\citep{bhh+03}, was clearly detected, and tight constraints on our
local motion with respect to this background were made with these
measurements. Analysis of the X-ray background at intermediate
redshifts ($z \la 4$) \citep{bck02} and the matter distribution at
local distances (the Local Group), probed at infrared wavelengths
using data from the {\it Infrared Astronomical Satellite} (IRAS)
\citep{wlf97}, have resulted in dipole detections as well. A marginal
detection was also reported by \citet{bw02} using radio data from the
NRAO VLA Sky Survey (NVSS) \citep{ccg+98}. In Section 4.2, I discuss
this radio survey specifically in the context of a dipole measurement.
\citet{bll+98} refer to a large number of previous dipole
investigations using infrared galaxy catalogs
\citep[e.g.,][]{md86,ywr86,vs87,hlm87}, optical catalogs
\citep[e.g.,][]{l87,lrl88,p88}, IRAS redshift surveys
\citep[e.g.,][]{rls+90,syd+92,wlf97}, and optical redshifts
\citep{llb89,h93}.  More details about inhomogeneities on large scales
in different kinds of surveys can be found in \citet{l00}, where an
extensive discussion is presented.

An isotropically distributed population of radio sources is predicted
to exist at cosmological redshifts in an isotropically expanding
universe. The extragalactic radio source population is observed to
have a median redshift of $z \sim 1$ \citep[e.g.,][]{lwl97}, 
a much later epoch than the cosmic microwave
background.  Assuming the radio source distribution is intrinsically
isotropic, our local motion with respect to the rest frame of this
population would produce a dipole anisotropy in the sky distribution
of radio sources.  Large-scale radio survey catalogs covering a large
fraction of the sky can in principle be used to detect this dipole
anisotropy and subsequently constrain the magnitude and direction of
our local motion with respect to this frame.

In this paper, I first describe the dipole anisotropy in the sky
distribution of radio sources that would be introduced by our motion
with respect to an isotropic radio source population. I then describe
the corresponding dipole signal that would be seen in the distribution
of discrete radio survey source counts. I describe a method to detect
this signal in radio survey catalogs, with particular attention to the
statistical significance of a possible detection.  Finally, I consider
both existing and proposed future radio surveys and whether a velocity
dipole signal would be detectable with statistical confidence using
this method given the number of discrete source counts present in
these surveys. In this analysis, I consider only the most optimistic
case in which there are no survey calibration problems, complications
from the masking of certain survey regions, or uneven survey coverage.

\section{The Effect of Motion on the Observed Radio Source Distribution}

Motion with respect to the rest frame of an isotropic distribution of
radio sources introduces two effects that change the observed source
number density on the sky as a function of sky position \citep{eb84,
c88}. The first effect is a boosting of the flux from sources located
toward the direction of motion. This depends on the observer speed and
the source spectral index. This Doppler effect changes the number of
detectable sources above a given flux threshold in a radio survey. The
second effect, relativistic beaming, changes the apparent solid angle
on the celestial sphere as seen by the moving observer. This in turn
alters the observed source number density as a function of sky
position.  For $N$ radio sources that are isotropically distributed
across the full $4\pi$ sr of sky, the source number density is
$\sigma_{0} = N/4\pi$, a constant. The density variation from the
observer's motion will depend on the angle $\theta$ measured from the
direction of motion (defined as $\theta = 0$) so that $\sigma(\theta)
= dN/d\Omega$.  Azimuthal symmetry in the situation removes any
dependence on the second angle $\phi$.

By combining the two effects described above, an expression for the
source number density of radio sources seen by the moving observer has
been presented by \citet{eb84}. For small observer speeds ($\beta
\equiv v/c \ll 1$) in which only first-order terms in $\beta$ are
retained, the source number density as a function of $\theta$ is shown
to be:

\begin{equation} 
\sigma(\theta) = \{1 + [ 2 + (\gamma-1)(1+\alpha)] \beta \cos \theta
\} \sigma_{0}
\label{eqn-neteffect}
\end{equation} 

which is a dipole anisotropy.  In this expression, $\gamma$ is the
power-law index for the number of extragalactic radio sources per flux
interval ($dN/dS \sim S^{-\gamma}$), and $\alpha$ is the mean radio
spectral index of the source population (defined here according to $S
\sim \nu^{-\alpha}$).  Following this, we can write:
 
\begin{equation} 
\sigma(\theta) = (1 + A \cos \theta) \sigma_{0},
\end{equation} 

where the amplitude $A = [2 + (\gamma-1)(1+\alpha)]\beta$. This
expression (with some variation in nomenclature convention) has been
used in the analysis of radio source counts by other authors
\citep[e.g.][]{c88, bll+98, bw02}. The mean values of the radio
spectral index and power-law index for the extragalactic radio source
population are estimated to be $\alpha \sim 0.75$, and $\gamma \sim
2$, respectively (e.g. Blake \& Wall 2002)\nocite{bw02}, and I use
these values in the analysis throughout the rest of the paper.

\section{Detecting the Dipole Anisotropy in Radio Survey Catalogs}

In this section, I describe a method for detecting the dipole
anisotropy in radio survey catalogs, and I calculate the minimum
number of discrete catalog sources required to detect the anisotropy
with statistical significance.  For the calculation, I assume complete
sky coverage (4$\pi$ sr) in which the survey has no masked,
incomplete, or invalid regions.  Perfect survey flux calibration is
also assumed.  These conditions, of course, do not apply in real
surveys, and I briefly address (but do not solve) the complications
posed by the presence of these effects, which can be quite
significant. Since survey masking and flux thresholds will reduce the
number of usable source counts in a survey, I do consider the effect
that this reduction has on the dipole detectability in two of the
surveys that I discuss.

\subsection{A Method for Detecting the Dipole Anisotropy}

A dipole vector $\vec{D}$ can be measured for a distribution of
discrete sources by computing an evenly weighted sum over all $N$
source counts, where each source contributes a three-dimensional unit
vector $\hat{r}$ to the sum (e.g., Baleisis et
al.~1998\nocite{bll+98}).\footnote{There is abundant previous
literature on the use of directional statistics and spherical data and
coordinates (like those employed here) as applied across a number of
scientific disciplines. Extensive discussions can be found in books by
\cite{b81}, \citet{fle93}, and \citet{mj99} in which applications of
directional statistics in a diverse range of subjects (e.g., biology,
meteorology, psychology, earth sciences) are presented.}  The
direction of $\hat{r}$ is determined by the position of the source on
the sky as seen by the observer at the center of the celestial
sphere. The sum is then:

\begin{equation}
\vec{D} = \sum_{i=1}^{N} \hat{r}_{i} 
\end{equation} 

In the case of an ideal survey with perfect flux calibration, complete
sky coverage, and in the limit of large $N$, all components of the
vector sum would cancel except for a component along the direction of
motion ($\theta=0$), corresponding to the $\hat{z}$-axis.

In this case, the sum can be written as the integral of the number
density over the celestial sphere:

\begin{equation}
\vec{D} = \hat{z} \int_{\phi = 0}^{\phi = 2\pi}
\int_{\theta=0}^{\theta=\pi} \sigma(\theta) \cos \theta \sin \theta
d\theta d\phi
\end{equation} 

The additional $\cos \theta$ term is present here since we are only
concerned with the $\hat{z}$-component of the integral. Replacing
$\sigma(\theta)$ in the expression with Equation 2 gives:

\begin{equation}
\vec{D} = \hat{z} \int_{\phi = 0}^{\phi = 2\pi} d\phi 
\int_{\theta=0}^{\theta=\pi} \sigma_{0} (1 + A \cos \theta) 
\cos \theta \sin \theta d\theta
\end{equation}

which, when evaluated, gives:

\begin{equation}
\vec{D} = \frac{4\pi}{3} A \sigma_{0} \hat{z} 
\end{equation} 

Using $\sigma_{0} = N/4\pi$ and $A = [2 + (\gamma-1)(1+\alpha)]\beta$,
the magnitude of the measured dipole can be written as:

\begin{equation} 
|\vec{D}| = D = \frac{1}{3} [2 + (\gamma-1)(1+\alpha)] \beta N
\end{equation} 

This is the dipole amplitude $D$ one would measure for an ideal survey
of $N$ radio sources with complete sky coverage if the observer were
moving with speed $v = \beta c$ (and $v \ll c$).

\subsection{The Statistical Significance of a Dipole Detection}

For a finite number of source counts $N$, there will be an uncertainty
in the measured dipole $\vec{D}$ arising from shot noise, and this
uncertainty determines the significance with which $\vec{D}$ can be
measured.  We can imagine the uncertainty as being represented by a
three-dimensional spherical probability cloud in direction space which
represents possible measured values of $\vec{D}$. This cloud will have
some characteristic radial size, and it will be centered on the tip of
the actual measured dipole $\vec{D}$.  The cloud is spherical since
there is an equal probability of the deviation occurring in any
three-dimensional direction.  This probability cloud can be used to
estimate the uncertainty in the measured speed of the observer
$\beta$, and exclude the zero-motion case ($\beta=0$) at some
confidence level (if an intrinsically isotropic distribution of radio
sources is assumed).  The direction of motion can also be determined,
with statistical uncertainties, by considering the direction of
$\vec{D}$ and the angular size that the probability cloud subtends as
seen from the origin (by the observer). For the test of whether the
zero-motion case is excluded with statistical significance, it is
functionally more useful to re-center the probability cloud at the
origin instead of centering it at the tip of $\vec{D}$. This is done
for the discussion below.

The characteristic size $\rho$ of the probability cloud is determined
by a random walk of $N$ steps in three dimensions (corresponding to
the $N$ sources in the vector sum that produces $\vec{D}$).  If the
magnitude of each step is 1 unit (corresponding to the size of the
unit vector), then we must consider the average magnitude of these
steps as projected onto an axis.  The actual step sizes along this
axis will range from $-1$ to $+1$ with a probability weighting that is
determined by the solid angle area from which the given projected
value comes.  The average projected step size $s$ on the axis is
computed from this to be $s = 1/\sqrt{3}$.

For large $N$, the probability distribution of the magnitude of the
final radial displacement is determined by the three one-dimensional
Gaussian distributions in Cartesian coordinates which represent the
three directional degrees of freedom.  The probability $p(r)$ that the
displacement occurs at a radial distance $r$ from the starting point
(the origin) is:

\begin{equation}
p(r) = \frac{1}{\rho^{3} (2 \pi)^{3/2}} 4 \pi r^{2} \exp(-r^{2} / 2 \rho^{2})
\end{equation} 

This is obtained by multiplying the three one-dimensional Gaussian
probability distributions and including an additional weighting factor
$4 \pi r^{2}$ to account for the increasing spherical volume element
as $r$ increases.

The characteristic size $\rho$ is determined by the random walk
according to:

\begin{equation}
\rho = s \sqrt{N} = \sqrt{ \frac{N}{3} } 
\end{equation} 

Using this, the probability $p(r)$ can be re-written in terms of $N$
as:\footnote{This result is not new, and the same expression can be
derived differently by computing the $\chi^{2}$ probability
distribution with three degrees of freedom,
$\chi_{3}^{2}(3r^{2}/N)$. Introducing a change in variables with a
Jacobian and employing a substitution eventually leads to a result
that is identical to Equation \ref{eqn-10}.}

\begin{equation}
p(r) = \left( \frac{54}{\pi N^{3}} \right)^{1/2} r^{2} \exp(-3r^{2}/2N)
\label{eqn-10}
\end{equation}

\begin{figure}
\centerline{\psfig{figure=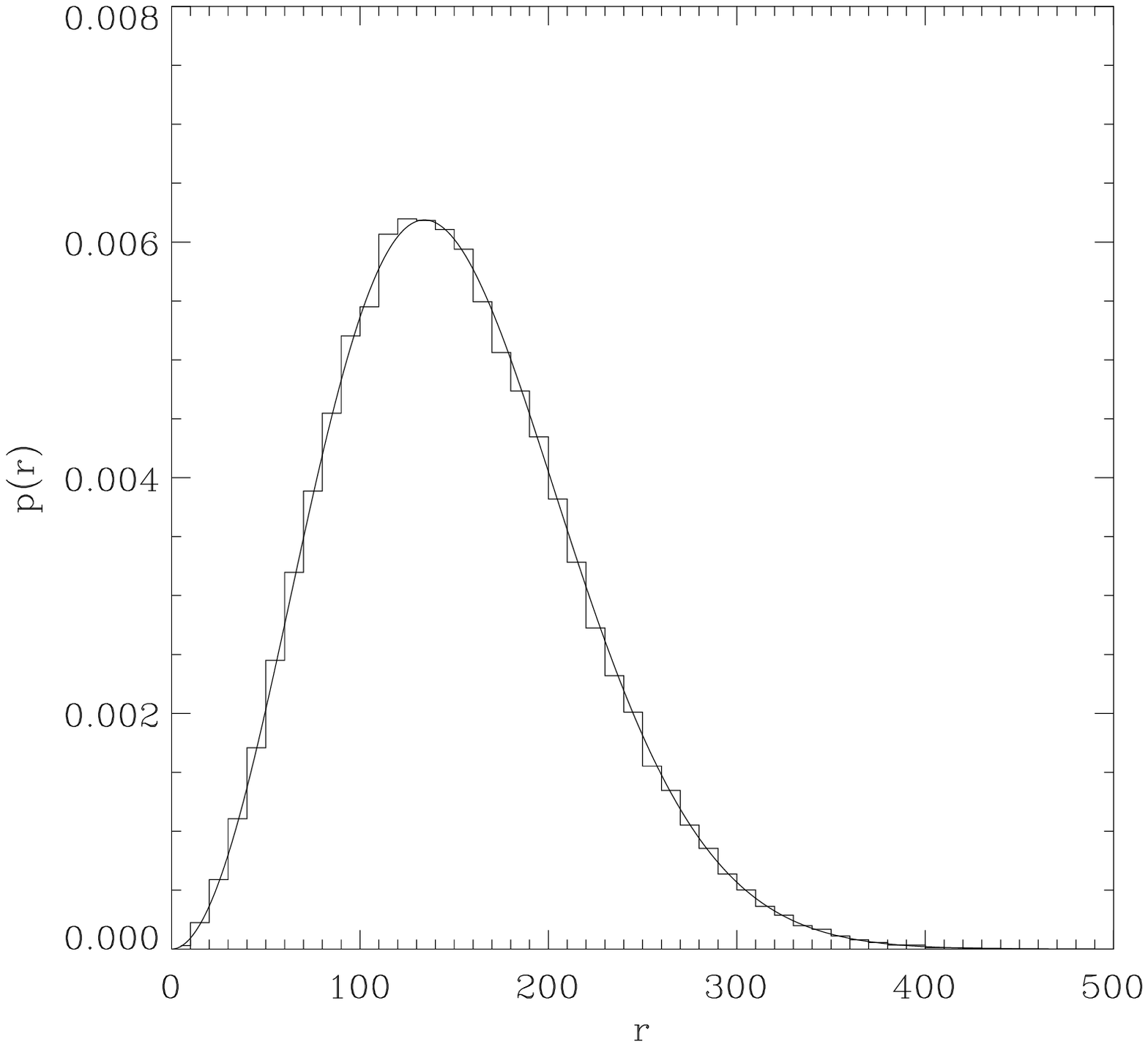,width=7in}}
\caption{Histogram of the magnitude of the displacement vectors $D$
calculated from 10$^{5}$ survey simulations, each of which had $2.7
\times 10^{4}$ sources isotropically distributed across the sky (see
also Figure \ref{fig-ball}, from which this histogram was
produced). Also plotted is the probability distribution $p(r)$ from
Equation \ref{eqn-10} as a function of radial displacement $r$, with
$N = 2.7 \times 10^{4}$ used in the probability expression. The
histogram has been normalized in the plot to have a total area equal
to one for comparison with $p(r)$. The match between the curve and the
histogram supports the use of this probability distribution in our
analysis.\label{fig-hist}}
\end{figure}

Figure \ref{fig-hist} shows the probability distribution $p(r)$ from
Equation \ref{eqn-10} plotted against a histogram of the magnitude of
the displacement vectors $D$ that were calculated from 10$^{5}$
simulated surveys. Each simulated survey had a random isotropic
distribution of $2.7 \times 10^{4}$ sources (which is the number of
sources selected for the analysis and discussion of the combined
87GB/PMN survey below; see also Table \ref{tbl-1}).  One of the
simulated surveys is shown in Figure \ref{fig-simuplot2} and is
discussed in more detail below.  The histogram in Figure
\ref{fig-hist} was produced from the set of displacement vectors shown
in Figure \ref{fig-ball} (this is discussed below), and it has been
normalized in Figure \ref{fig-hist} to have a total area equal to one
for comparison with the probability curve. The match between the curve
and the histogram supports the use of this probability distribution in
the subsequent calculations and analysis.

We can use Equation \ref{eqn-10} to estimate the likelihood of
randomly measuring a dipole with a magnitude greater than or equal to
$D$ in the zero-motion, isotropic case. We integrate the probability
function $p(r)$ from $r = D$ to an upper limit of $r = N$ (which is
the maximum possible dipole value; in the limit of large $N$, the
value of this upper limit approaches infinity).  The integration is
purely radial since the angular dependence drops out for the spherical
cloud.  The result of the integration, $p(r > D)$, which ranges
between 0 and 1, is the probability-weighted volume outside of the
radius $D$.  This represents the likelihood of randomly measuring a
dipole of magnitude $D$ or greater from an isotropic source
distribution.  Conversely, we can use the measured dipole magnitude
$D$ to exclude the zero-motion case ($\beta=0$) at a confidence level
determined by $1 - p(r > D)$ (e.g., if $1-p$ = 68\%, then the
zero-motion case would be excluded at the 1$\sigma$ level).

We can also determine an uncertainty in the magnitude of the measured
dipole $D$ using similar reasoning. We surround the measured vector
$\vec{D}$ with the spherical probability cloud centered at its tip. We
pick the confidence level of interest (1$\sigma$, 2$\sigma$, etc.; let
us call it $n \sigma$) from the corresponding percentage level of
exclusion. Using this, we solve for the corresponding radial distance
$r_{n}$ in the probability cloud for which the probability-weighted
volume within $r_{n}$ is the aforementioned percentage of the total
probability-weighted volume. This is done by evaluating the integral
for $p(r)$ using a lower limit ($r_{n}$) which will give the
appropriate percentage level of exclusion.  Once $r_{n}$ has been
determined in this way, this defines an $n \sigma$ error sphere
centered on the tip of $\vec{D}$.  Since the uncertainty cloud is
spherical, the projection of the cloud onto the $\hat{D}$ axis gives
the magnitude $r_{n}$, which is the $n \sigma$ uncertainty in $D$.
The measured dipole magnitude is then $D \pm r_{n}$ (at the $n \sigma$
confidence level).  Figure \ref{fig-1} shows detection confidence
levels of dipole detections for calculations of $D$ and $r_{n}$ from
several idealized surveys with different numbers of source counts $N$
and different assumed values for the dipole velocity $\beta$.

This error sphere can also be used to determine an uncertainty in the
direction of $\vec{D}$. The projection of the error sphere onto the
celestial sphere as seen by the observer at the center defines a
circularly symmetric region on the sky centered on the dipole
direction $\hat{D}$. Sky directions lying outside the circular region
are statistically excluded at the $n\sigma$ level.  Let the dipole
direction $\hat{D}$ correspond to $\theta = 0$ again, and let the
$n\sigma$ angular uncertainty in the dipole direction be $\delta
\theta_{n}$ (this is the angular radius of the projected circular
region).  $\delta \theta_{n}$ is determined by the angle between
$\hat{D}$ and a vector which starts at the origin and is tangent to
the error sphere of radius $r_{n}$. This is calculated by:

\begin{equation}
\delta \theta_{n} = \arcsin (r_{n} / D)
\end{equation}

for $r_{n} < D$. No statistically significant direction constraint can
be made for the case $r_{n} \ge D$. This uncertainty angle $\delta
\theta_{n}$ is determined by the number of sources $N$ in the survey
and the measured value of $D$ (which in turn depends on $N$ and the
observer's velocity $\beta$). Figure \ref{fig-2} shows calculations of
$\delta \theta_{n}$ for an assumed dipole velocity of 370 km s$^{-1}$
as a function of the number of survey source counts $N$.

\subsection{Incomplete Survey Sky Coverage and Imperfect Flux Calibration}

For real radio surveys which do not have complete sky coverage, an
artificial dipole would be measured when the sum of the unit vectors
$\hat{r}$ is taken over the $N$ observed sources.  This dipole will be
biased toward directions which have more complete survey coverage.  To
deal with this incompleteness when analyzing real radio survey data,
one usually defines a mask of survey regions on the celestial sphere
to exclude in the analysis. The vector sum is then taken over the
distribution of sources in the unmasked, valid regions to obtain a
measured dipole $\vec{D}$. The center of the uncertainty probability
cloud (described above) will be offset from the origin (corresponding
to a magnitude and direction) by an amount determined by the geometry
of the unmasked region.  In principle, one can simply subtract this
offset vector from all other vectors in the analysis (such as the
measured $\vec{D}$) so that everything (including the distribution of
points in the probability cloud) is re-centered to the origin. One
would then proceed with the analysis as described above.

\begin{figure}
\centerline{\psfig{figure=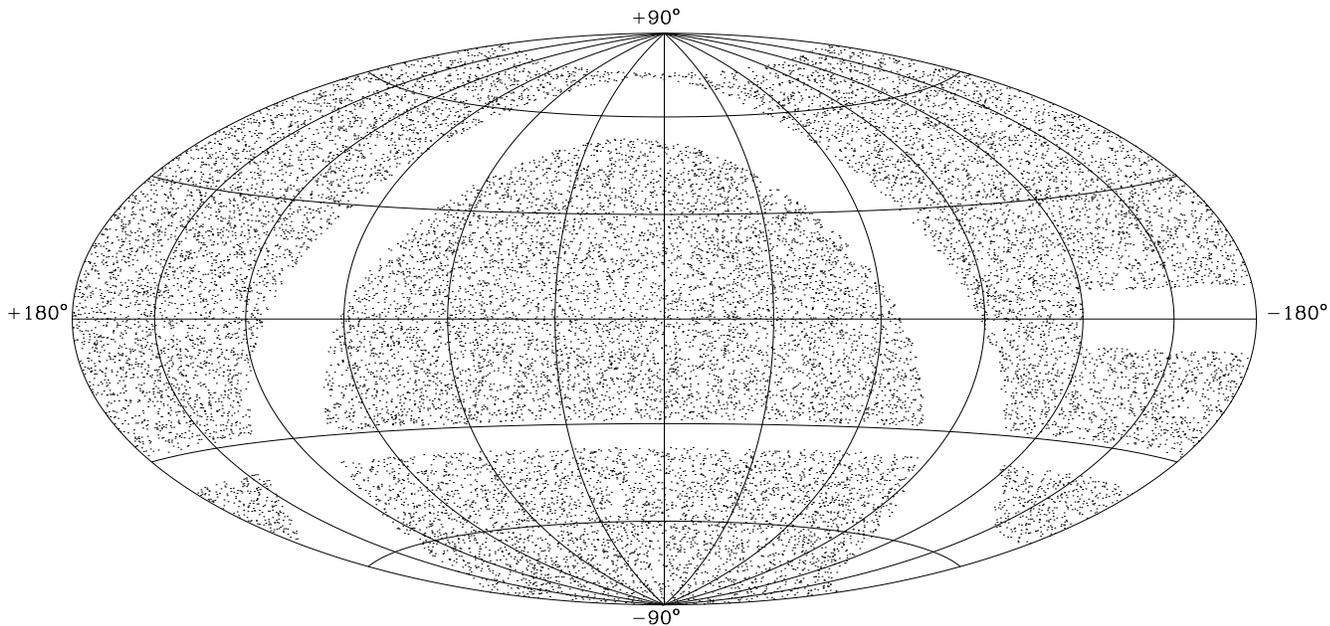,width=7in,angle=90}}
\caption{Aitoff projection plot in equatorial coordinates of $2.7
\times 10^{4}$ simulated radio sources distributed isotropically
across the unmasked regions of the sky. The masked regions shown here
are very close to the ones used by \citet{lwl97} and \citet{bll+98} in
their analysis of the combined 87GB/PMN survey (see also the
discussion of these surveys in the text and Table \ref{tbl-1}).  The
$2.7 \times 10^{4}$ sources in this simulated survey correspond to the
number of sources in the 87GB/PMN survey in the unmasked regions which
have flux densities between 50 and 100 mJy. See Figure 1 of
\citet{bll+98} for a qualitative comparison of this simulated source
distribution with the actual source
distribution.\label{fig-simuplot2}}
\end{figure}

\begin{figure}
\centerline{\psfig{figure=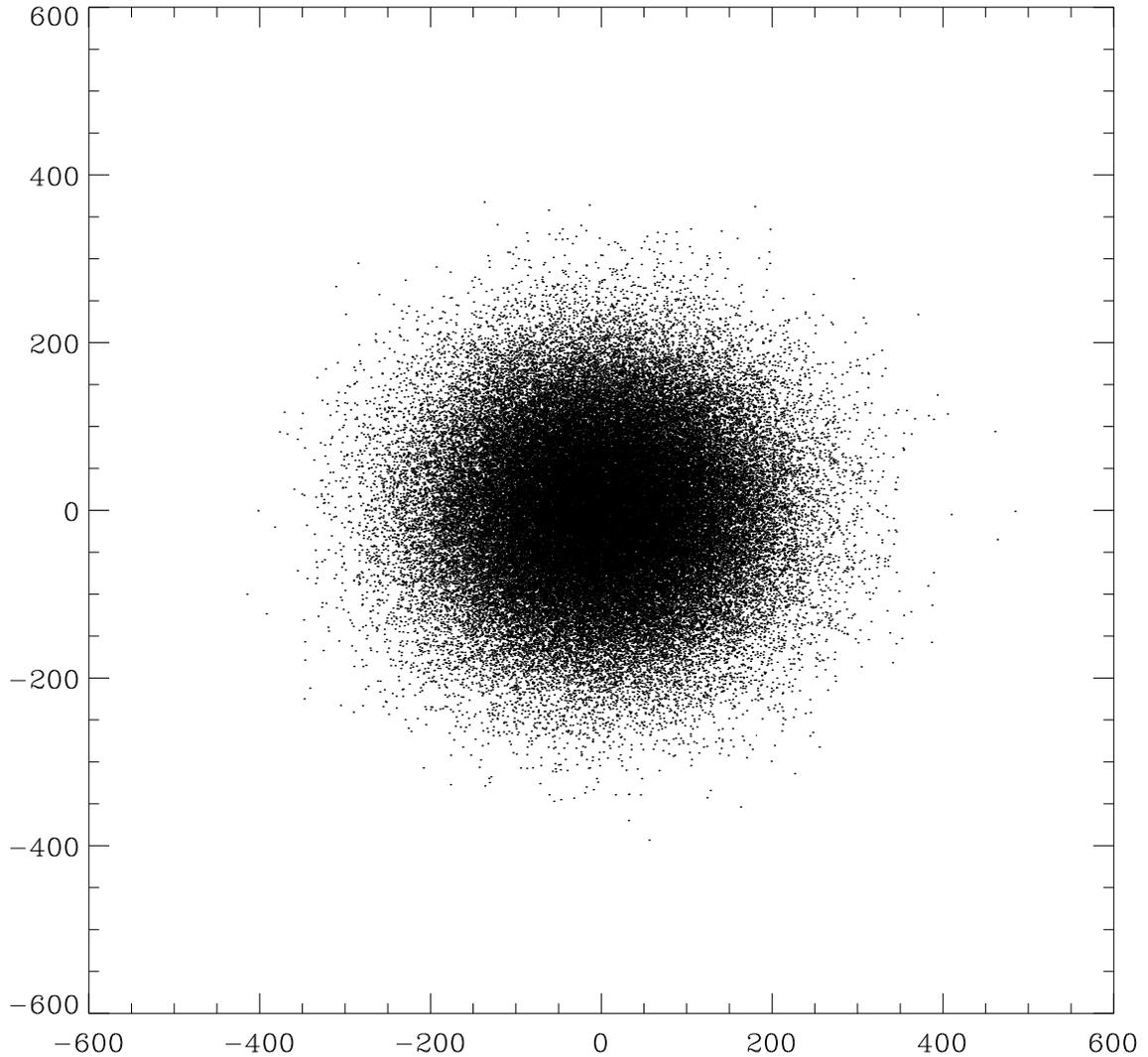,width=6.5in}}
\caption{Two dimensional projection plot of the set of displacement
vectors $\vec{D}$ in three dimensions that were produced from 10$^{5}$
simulations of isotropic surveys, each of which had $2.7 \times
10^{4}$ sources. One of these simulated surveys is shown in Figure
\ref{fig-simuplot2}. The distribution shown here was produced after
the mean of the distribution was subtracted from each displacement
vector to correct for the vector offset introduced from survey masking
effects. This three-dimensional distribution was used to produce the
histogram of radial displacements shown in Figure
\ref{fig-hist}.\label{fig-ball}}
\end{figure}

However, even without any problems introduced by the flux calibration,
the shapes of the regions to be masked in a real survey are generally
complicated. For example, sources in the Galactic plane and in certain
declination bands may not be appropriate to include in the analysis
(e.g., see the data masking used in the analysis of radio surveys by
\citet{lwl97}, \citet{bll+98}, and \citet{bw02}; see also Figure
\ref{fig-simuplot2}).  Also, the finite number of sources introduces a
shot noise uncertainty in the offset vector.  In the absence of
calibration issues, the most straightforward course of action would be
to simulate the survey by randomly assigning positions to $N$ sources
drawn from an isotropic distribution within the unmasked region of the
sky and compute the sum of source unit vectors for many of these
simulations.  Figure \ref{fig-simuplot2} shows an Aitoff projection in
equatorial coordinates of a single simulated survey of $2.7 \times
10^{4}$ sources distributed isotropically over the unmasked regions of
the sky (cf. Figure 1 of \citet{bll+98} and their masked regions,
which were used as a guide here). The set of resulting vector
displacements from the survey simulations will produce a probability
cloud in which the mean position defines the vector to be
subtracted. Figure \ref{fig-ball} shows the vector displacements
produced from 10$^{5}$ simulated isotropic surveys like the one shown
in Figure \ref{fig-simuplot2}. In this plot, the mean vector has
already been subtracted so that the cloud is centered at the origin,
and the resulting displacements $\vec{D}$ have been projected onto a
two-dimensional plane.  This distribution of displacements in Figure
\ref{fig-ball} is the probability cloud that was used to produce the
histogram shown in Figure \ref{fig-hist}.

Source clustering at local distances can also affect the analysis of
radio surveys (e.g., the NVSS Survey; see discussions by Boughn \&
Crittenden 2002\nocite{bc02}, Blake and Wall 2002\nocite{bw02}, and
also below). In this paper I make the assumption that the intrinsic
distribution of sources is homogeneous and that there is no
significant clustering on large scales beyond the local mass
distribution. I assume that local sources would be excised prior to
analysis.

Also, real radio surveys do not have perfect flux calibration, and
this can bias the source count distribution in certain sky regions,
thereby producing an artificial dipole anisotropy.  The effects of
survey calibration errors in combination with the masking of survey
regions can be quite severe, and dealing with these complex effects
has been much discussed in the literature (e.g., \citet{hgn+02} for
the analysis of the cosmic microwave background, and \citet{ht04} and
\citet{sth+08} for large-scale galaxy catalogs, such as the Sloan
Digital Sky Survey \citep{yaa+00}). In the analysis presented here, I
ignore all such calibration effects and consider only the raw source
counts $N$ and whether a dipole would be detectable in the idealized
case where the surveys and instruments are perfect.

\section{Prospects for a Dipole Detection in Existing and Proposed Future 
Radio Surveys}

Large-scale radio surveys currently exist in which the method outlined
here for detecting the velocity dipole can be applied. However, the
statistical significance of such a detection depends on whether these
surveys have a sufficient number of sources, even if the flux
calibration were perfect. Next-generation radio instruments that are
in various stages of planning and development will also be used to
conduct large-scale radio surveys in the future.  Below I consider
existing and possible future radio surveys in the order of increasing
number of source counts, and I discuss the possibility of detecting
the dipole in each of the surveys in turn, assuming the most
optimistic scenario (i.e., perfect flux calibration).

\subsection{The 87 Green Bank and Parkes-MIT-NRAO Surveys} 

The 87 Green Bank (87GB) \citep{gc91} and Parkes-MIT-NRAO (PMN)
surveys \citep{gwb+94, wgb+94, wgh+96} were conducted at 4.85 GHz and
covered the northern and southern celestial sky, respectively.
\citet{bll+98} and \citet{lwl97} give concise descriptions and
summaries of these surveys in the context of large-scale structure and
the velocity dipole effect, and they refer to a number of previous
attempts to detect the velocity dipole (see also the references in
Section 1). \citet{bll+98} have conducted an analysis of whether the
87GB and PMN surveys could be used for detecting the dipole using a
method similar to the one described here (in their case, they
considered the measured dipole magnitude relative to the shot noise
term but did not do a more extensive analysis of possible detection
significance).

\begin{figure}
\centerline{\psfig{figure=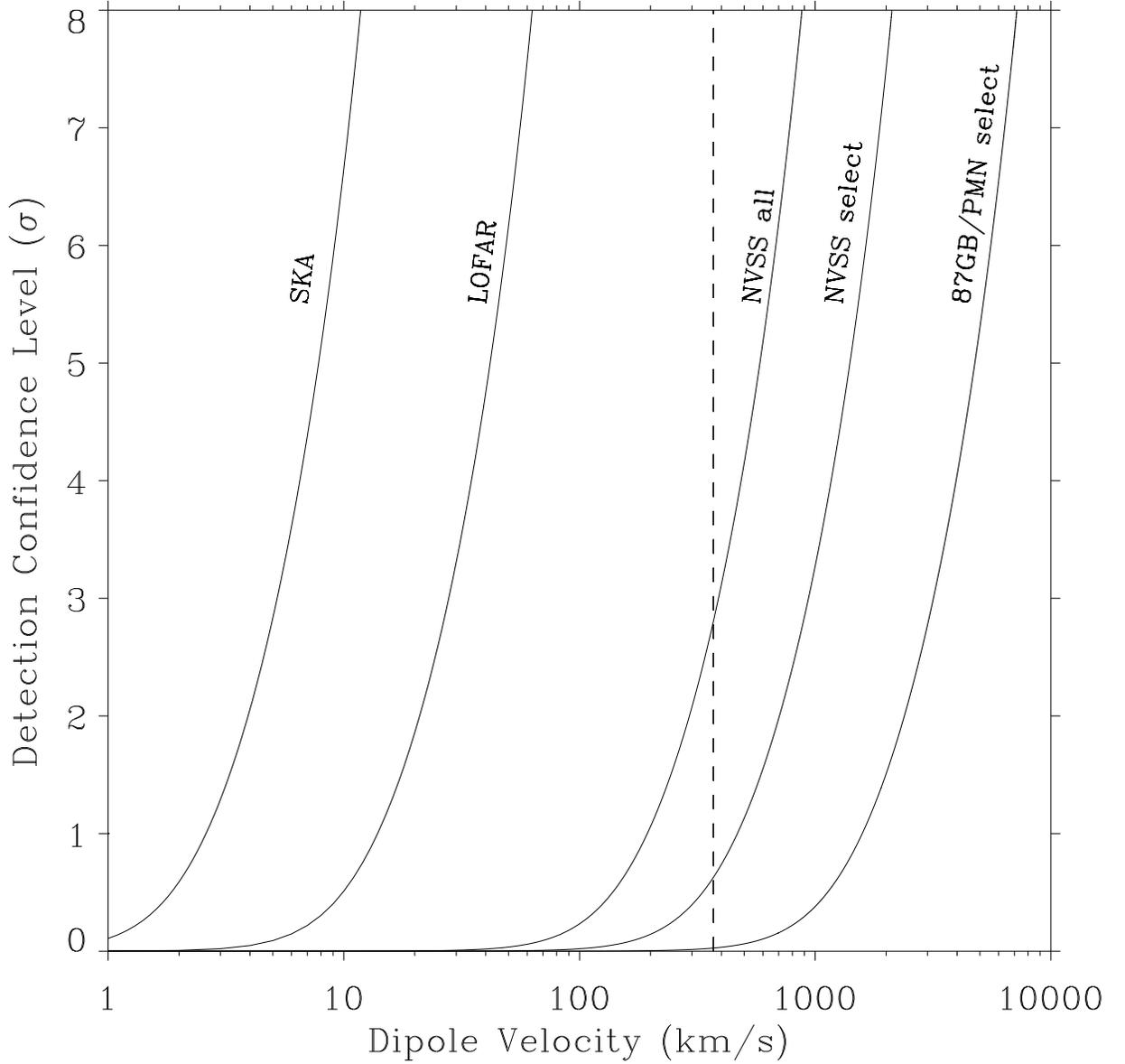,width=7in}} 
\caption{Dipole detection confidence level vs. dipole velocity for
several existing and proposed future large-scale radio surveys. The
solid curves represent the various surveys with different numbers of
sources and are labeled (see the text and Table \ref{tbl-1} for survey
descriptions).  In this plot, each survey is assumed to have perfect
flux calibration and no contamination from local sources,
corresponding to the most optimistic detection case possible.  The
expected dipole velocity of 370 km s$^{-1}$ is indicated by the dashed
vertical line.  For this velocity, a minimum of $2.0 \times 10^{6}$,
$3.1 \times 10^{6}$, and $4.5 \times10^{6}$ survey sources would be
required for 3$\sigma$, 4$\sigma$, and 5$\sigma$ dipole detections,
respectively.\label{fig-1}}
\end{figure}

\begin{figure}
\centerline{\psfig{figure=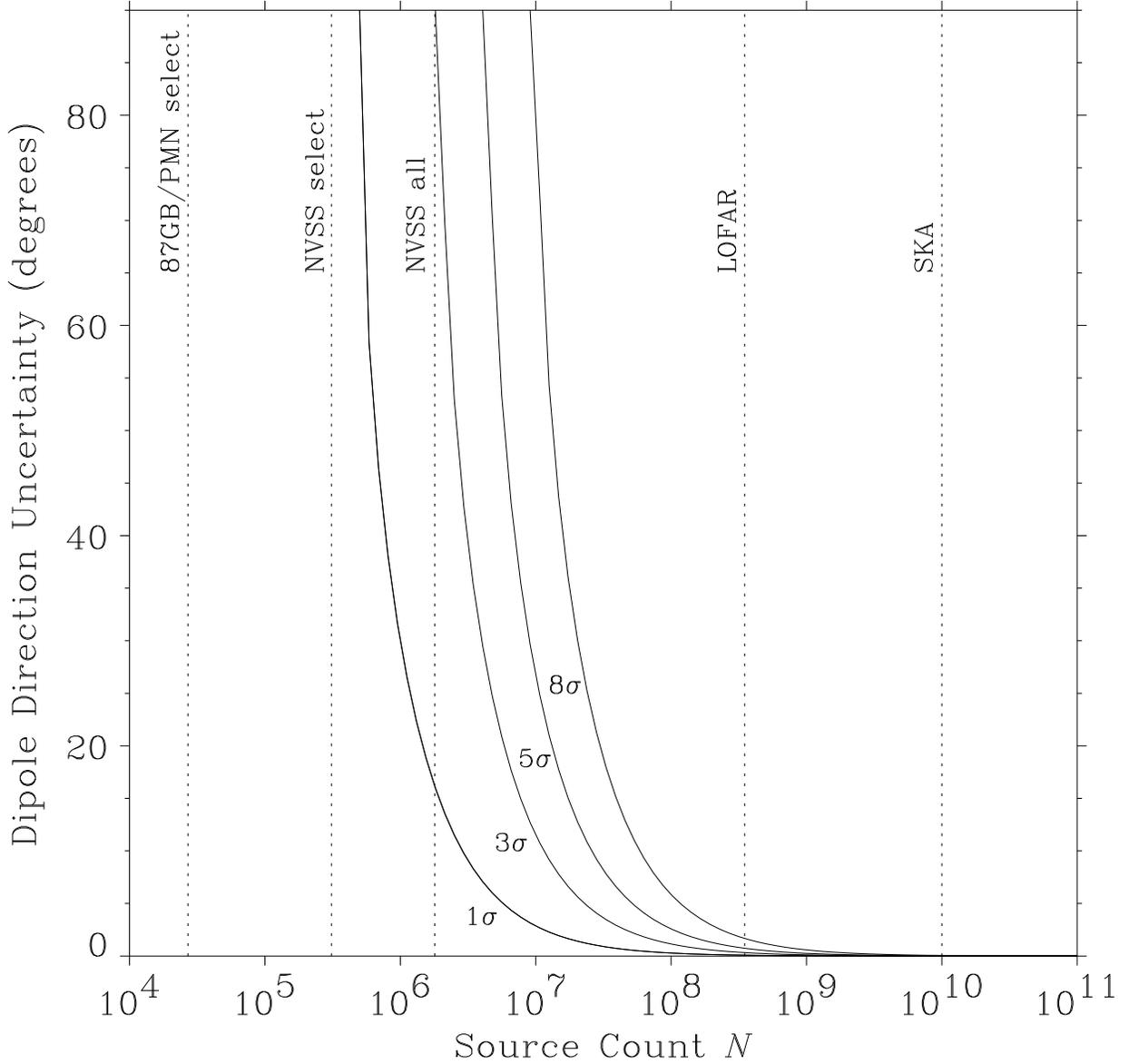,width=7in}}
\caption{Uncertainty in the measured dipole direction ($\delta
\theta$) vs. number of survey source counts ($N$) for several existing
and proposed future large-scale radio surveys (see the text and Table
\ref{tbl-1}). In this plot, a dipole velocity of 370 km s$^{-1}$ is
assumed as is perfect flux calibration and no contamination from local
sources (corresponding to the most optimistic detection case
possible).  The dotted vertical lines represent the various surveys
with different numbers of sources and are labeled.  From left to
right, the solid curves represent 1$\sigma$, 3$\sigma$, 5$\sigma$, and
8$\sigma$ constraints on the dipole direction uncertainty that are
possible with this detection method.\label{fig-2}}
\end{figure}

In their analysis, \citet{bll+98} excised certain incomplete or
unreliable survey regions identified by \citet{lwl97}, including sky
regions within 10$^{\circ}$ of the Galactic plane, to eliminate
Galactic sources, and they imposed a minimum flux density cutoff of 50
mJy for sources to be considered. This latter selection reduced
declination-dependent number density variations.  After performing a
similar excision (but retaining only sources between 50 and 100 mJy),
I find that the selected sample consists of $\sim 2.7 \times 10^{4}$
sources covering $\sim 70$\% of the celestial sphere (see Table
\ref{tbl-1}).  Figure \ref{fig-simuplot2} shows similar masking used
in a simulated survey of $2.7 \times 10^{4}$ isotropically distributed
sources.  As was also found by \citet{bll+98}, I find that this is not
an adequate number of source counts for a statistically significant
dipole detection given the expected local velocity of $v \sim 370$ km
s$^{-1}$, even if these surveys had perfect flux calibration. As seen
in Figure \ref{fig-1}, an observer velocity that is an order of
magnitude greater than this would be required for even a 3$\sigma$
detection. Figure \ref{fig-2} shows that for the expected velocity, no
direction constraint could be made with this sample at any
significance level.  The shot noise analysis conducted by
\citet{bll+98} showed that at least $\sim 4 \times 10^{5}$ galaxies
over the sky would be needed to detect the velocity dipole at the same
level as the shot noise (i.e., a $1\sigma$ detection). Our results are
consistent with theirs and suggest a comparable number of sources
would be needed for a $1\sigma$ detection (see Figure \ref{fig-1} and
Table \ref{tbl-1}, particularly the NVSS select survey that has $\sim
3.1 \times 10^{5}$ sources). This result does not account for the
dipole introduced from large-scale structure, which complicates the
analysis further. Thus, the 87GB and PMN surveys are not adequate for
detecting a velocity dipole anisotropy using this method, regardless
of whether slightly different flux density selection criteria are used
in the data selection.

\subsection{The NRAO VLA Sky Survey}

The NVSS is a large-scale radio survey that was conducted with the
Very Large Array at 1.4 GHz \citep{ccg+98}.  The total number of
sources in the NVSS is $\sim 1.8 \times 10^{6}$, and the survey is
estimated to be 99\% complete down to an integrated flux density of
3.5 mJy.  The survey covers 82\% of the celestial sphere,
corresponding to a declination range $\delta > -40^{\circ}$, and the
majority of the sources in the survey are believed to be at
cosmological distances.

\citet{bw02} have searched the NVSS for a dipole anisotropy, and we
follow their work to determine which survey regions and flux ranges
could be considered reliable for the dipole search described
here. Regions within $15^{\circ}$ of the Galactic plane were masked by
\citet{bw02} to eliminate Galactic sources in their dipole analysis.
They also eliminated the ``clustering dipole'' (from the Local
Supercluster), which has its own dipole component, thereby ensuring
that only cosmological sources were used in the analysis. This was
done by eliminating radio sources within 30$''$ of nearby galaxies
known from several catalogs.  Although the claimed 99\% completeness
level of the NVSS is 3.5 mJy \citep{ccg+98}, there are significant
declination effects evident at this flux level in which the number
density fluctuates by a few percent from the mean. \citet{bw02} use a
minimum flux cutoff of 15 mJy for their analysis (see their Table 1),
which reduces the variations to less than 1\%. However, as described
above, these flux variations in combination with survey masking may
still introduce complications which can significantly contaminate a
dipole search.

After eliminating unreliable survey regions, \citet{bw02} retained a
selected sample of $\sim 3.1 \times 10^{5}$ sources above 15 mJy,
representing $\sim 20$\% of the initial sample (see Table
\ref{tbl-1}). This is not adequate for a statistically significant
dipole detection with the method presented here, even if flux
calibration were perfect in the survey. For an expected dipole
velocity of 370 km s$^{-1}$, the best possible dipole detection would
be slightly less than 1$\sigma$ (Figure \ref{fig-1}). The dipole
direction would also be unconstrained (Figure \ref{fig-2}). Given
these issues, I conclude that the NVSS cannot be used for detecting
the velocity dipole using this method. Including all of the sources in
the entire NVSS catalog ($N \sim 1.8 \times 10^{6}$) would yield a
marginal detection and constraint at best ($\sim 3\sigma$), but flux
calibration problems with the weakest sources in the survey prevent
this from being feasible.

\subsection{Next-generation Surveys with LOFAR and the SKA}

Future large-scale radio surveys will be conducted with
next-generation radio telescope facilities that are currently in
various stages of planning and development. Among these new
instruments are the Long Wavelength Array
(LWA)\footnote{http://lwa.unm.edu} \citep[e.g.,][]{t06}, the Murchison
Widefield Array (MWA)\footnote{http://www.haystack.mit.edu/mwa}
\citep[e.g.,][]{mbc+06}, and the Low Frequency Array
(LOFAR)\footnote{http://www.lofar.org}, all of which will observe the
sky at low radio frequencies (a few hundred MHz or less). Farther into
the future, an even more advanced radio facility, the Square Kilometer
Array (SKA)\footnote{http://www.skatelescope.org}, is being considered
for development.  Although the details of the surveys to be conducted
with these instruments are not yet firmly established, I focus on two
possible surveys that have been outlined for two of these instruments:
LOFAR and the SKA.

LOFAR is an advanced radio telescope array that is expected to operate
at low radio frequencies (30 to 240 MHz) and will have thousands of
antenna elements distributed over hundreds of kilometers
\citep[e.g.,][]{svk+07,fvd+07,fws+08}.  A survey using LOFAR to search
for gravitational lenses has been outlined by
Jackson\footnote{N. Jackson, LOFAR Memorandum Series \#4}.  In a
possible LOFAR survey, half of the celestial sky (2$\pi$ sr) would be
covered down to a limiting sensitivity of $\sim 0.7$ mJy at 151 MHz.
The total number of sources expected to be detected in such a survey
would be $\sim 3.5 \times 10^{8}$ (Table \ref{tbl-1}). Other recent
estimates of source counts from proposed LOFAR surveys at a variety of
wavelengths suggest a range of $\sim 10^{7}$ to $\sim 10^{9}$
detectable sources in $2 \pi$ sr of sky \citep[see, for example, the
presentation by][]{r07}. For the sake of simplicity, we use the survey
parameters outlined by Jackson for our analysis in which the expected
number of source counts falls near the middle of this range.

The SKA is a planned next-generation radio telescope facility which
will have vastly increased sensitivity for large-scale radio surveys.
Simulations of a large-scale 1.4 GHz radio survey with the SKA suggest
a limiting flux density of $\sim 0.1$ $\mu$Jy and a yield of $\sim
10^{9}$ or more sources per sr \citep{hej+99, hwc+00}. This
corresponds to $\sim 10^{10}$ sources for the entire celestial sphere
(Table \ref{tbl-1}).

Even with large gaps in sky coverage, the sheer number of sources in a
survey conducted with either LOFAR or the SKA would far exceed the
requirements for a statistically significant dipole detection with
this method. If flux calibration problems can be sufficiently
minimized or properly corrected, and if local source contamination can
be removed, it should be easy to detect the dipole with large
statistical significance if the dipole velocity is near the expected
value of $v \sim 370$ km s$^{-1}$ (see Figure \ref{fig-1}).  The
constraint on the dipole direction should also be very precise (Figure
\ref{fig-2}), easily constrained to within a few degrees or less in
both cases.

\section{Conclusions} 

I have described a method for detecting a velocity dipole anisotropy
in large-scale radio surveys and considered the feasibility of
detecting this dipole with statistical significance in existing and
proposed future large-scale surveys. This analysis does not account in
any way for the severe complications that arise from imperfect flux
calibration and masking effects in surveys, and therefore corresponds
to the most optimistic detection case possible. Neither the combined
87GB/PMN survey nor the NVSS has a sufficient number of sources to
detect the velocity dipole anisotropy with statistical significance
using this method, even if no calibration or sample bias effects were
present.  However, proposed large-scale radio surveys using
next-generation radio science instruments (e.g., LOFAR and the SKA)
are more promising: surveys with these instruments should easily have
enough source counts for a statistically significant dipole detection
and direction constraint if flux calibration problems and
contamination from local sources can be sufficiently reduced or
eliminated.

\acknowledgements 

I thank Elizabeth Praton for assistance with some of the derivations,
Nathan Keim for contributions to the simulation work, and Steve Boughn
and the referee John Ralston for helpful comments and insights which
have improved this work.


\begin{deluxetable}{lccc}
\tablecaption{Existing and Proposed Future Large-Scale Radio Surveys.\label{tbl-1}}
\tablewidth{0pt}
\tablehead{
\colhead{Radio Survey} &
\colhead{Observing Frequency} &
\colhead{Flux Density Limit} &
\colhead{$N$} \\
\colhead{} & 
\colhead{(GHz)} &
\colhead{(mJy)} &
\colhead{} 
}
\startdata
87GB/PMN select\tablenotemark{a} & 4.85                  & 50        & $2.7 \times 10^{4}$  \\
NVSS select\tablenotemark{b}     & 1.4                   & 15        & $3.1 \times 10^{5}$  \\
NVSS all\tablenotemark{c}        & 1.4                   & 3.5       & $1.8 \times 10^{6}$  \\
LOFAR\tablenotemark{d}           & 0.15\tablenotemark{f} & 0.7       & $3.5 \times 10^{8}$  \\
SKA\tablenotemark{e}             & 1.4\tablenotemark{f}  & $10^{-4}$ & $\sim 10^{10}$       
\enddata

\tablecomments{``Select'' refers to a selected sample of sources from
the survey deemed appropriate for the analysis.}

\tablenotetext{a}{Includes only sources with a flux density between 50
and 100 
mJy from the combined 87GB and PMN surveys. Several regions, including
regions within 10 degrees of the Galactic plane, were also excluded
\citep{lwl97, bll+98}.}

\tablenotetext{b}{Includes only sources with a flux density above 15
mJy from the NVSS.  Several regions, including regions within 15
degrees of the Galactic plane, were also excluded \citep{bw02}.}

\tablenotetext{c}{Includes all sources in the NVSS with a flux density
above 3.5 mJy.}

\tablenotetext{d}{Includes all sources in a possible LOFAR 
survey outlined by Jackson.}

\tablenotetext{e}{Includes all sources in a possible SKA survey.}

\tablenotetext{f}{Possible observing frequency.}

\end{deluxetable}


\end{document}